\documentclass[nofootinbib,preprint, prX]{revtex4-1} 
\usepackage{multido}
\usepackage{mathrsfs}
\usepackage{amsmath}
\usepackage{amssymb}
\usepackage{amsthm}
\usepackage{dsfont}
\usepackage{amsfonts}
\usepackage{fullpage}		
\usepackage{xspace}
\usepackage{graphicx}		
\usepackage{graphics}

\usepackage{hyperref}

\usepackage{ulem}

\newcommand{\pDeut}{\hat{\mathbf{\epsilon}}_{d}^{*}}
\newcommand{\vect}[1]{\boldsymbol{#1}}

\newcommand{\pGamma}{\hat{\mathbf{\epsilon}}_{\gamma}^{*}}

\newcommand{\pDeutm}[1]{\hat{\epsilon}_{d_{#1}}^{*}{}\!}
\newcommand{\pGammam}[1]{\hat{\epsilon}_{\gamma_{#1}}^{*}{}\!}

\newcommand{\comment}[1]{}

\newcommand{\EFT}{$\mathrm{EFT}_{\not{\pi}}$\xspace}
\newcommand{\LO}{\mathrm{LO}}
\newcommand{\NLO}{\mathrm{NLO}}

\makeatletter
\newcommand{\vast}{\bBigg@{3}}
\newcommand{\Vast}{\bBigg@{5}}
\makeatother

\newcommand{\LRd}{\overset{\leftrightarrow}{D}}

\newcommand{\wave}[3]{\ensuremath{{}^{#1}\mathrm{#2}_{#3}}\xspace}
\newcommand{\oneS}{\wave{1}{S}{0}}

\newcommand{\threeS}{\wave{3}{S}{1}}

\renewcommand{\emph}[1]{\textit{#1}}

\newcommand{\MeV}{\text{MeV}}

\begin{document}

\title{Energy dependence of the parity-violating asymmetry of circularly polarized photons in $d\vec{\gamma} \to np$ in pionless effective field theory}

\author{Jared Vanasse}
\email{jjv9@phy.duke.edu}
\affiliation{\emph{Department of Physics,\\
Duke University,\\
Durham, NC 27708}
}

\author{Matthias R.~Schindler}
\email{mschindl@mailbox.sc.edu}
\affiliation{\emph{Department of Physics and Astronomy,\\
University of South Carolina,\\
Columbia, SC 29208}
}

\date{\today}

\begin{abstract}
We calculate the energy dependence of the asymmetry in the cross sections for circularly polarized photons on an unpolarized deuteron target in $d\vec{\gamma} \to np$ in pionless effective field theory.  By matching the parity-violating low-energy constants to different sets of corresponding model parameters we obtain estimates for the asymmetry.  In addition we calculate two possible figures of merit for the asymmetry in order to assess the preferred photon energy at which to perform a possible future experiment at a high-intensity photon source.
\end{abstract}


\maketitle

\section{Introduction}

Weak interactions between quarks in the nucleon induce a parity-violating (PV) component in the interactions between nucleons.
This PV interaction is expected to be suppressed by about six to seven orders of magnitude ($G_{F}m_{\pi}^{2}\sim10^{-7}$) compared to the dominant parity-conserving (PC) component.
While weak interactions are well understood at the level of quarks, the nonperturbative nature of the strong interactions at low energies makes it difficult to derive their manifestation at the nucleon level.
At the same time, the short range of the weak interactions allows hadronic parity violation to be considered as a unique probe of nonperturbative strong interactions.
For reviews, see e.g.~Refs.~\cite{Adelberger:1985ik,RamseyMusolf:2006dz,Haxton:2013aca,Schindler:2013yua}.

Traditionally, experimental studies of hadronic parity violation have been focused on systems containing a larger number of nucleons, where PV effects can be enhanced by several orders of magnitude due to close-lying states of opposite parity, see e.g.~Ref.~\cite{Bowman:1989ci}. 
However, many-body effects significantly complicate the theoretical analysis in terms of nucleon-nucleon (NN) interactions. 
A number of PV observables in two-nucleon systems can in principle be used to constrain the PV part of the nucleon-nucleon interaction. 
In addition to proton-proton scattering \cite{Balzer:1980dn,Kistryn:1987tq,Eversheim:1991tg,Berdoz:2002sn},  a particularly prominent example is the photon angular asymmetry $A_\gamma$ in the capture of polarized neutrons on unpolarized protons, which in terms of the traditional meson exchange models provides information on the PV pion-nucleon coupling.
The NPDGamma experiment at Oak Ridge National Laboratory's Spallation Neutron Source is currently determining this asymmetry with the aim of significantly improving on previous results \cite{Gericke:2011zz}.

A second observable in this system is the circular photon polarization $P_\gamma$ in the capture of unpolarized neutrons on unpolarized protons, $np\to d\vec{\gamma}$. 
While both $A_{\gamma}$ and $P_{\gamma}$ involve neutron capture, the different polarizations result in two complementary and independent observables.
In particular, the component of the PV NN interaction giving the main contribution to $A_\gamma$ is highly suppressed in $P_\gamma$, which thus presents the opportunity to constrain different PV components.
The requirement of a high neutron flux and the difficulty of measuring the circular polarization of the outgoing photon make a measurement of $P_\gamma$ challenging. A previous experiment was able to put a bound on $P_\gamma$ which, however, is consistent with zero \cite{Knyazkov:1984zz}.

For identical kinematics, the polarization $P_\gamma$ is equal to the longitudinal asymmetry $A^\gamma_L$,
\begin{equation}
A^\gamma_L = \frac{\sigma_+ - \sigma_-}{\sigma_+ + \sigma_-},
\end{equation} 
in the time-reversed process $\vec{\gamma}d\to np$, where $\sigma_\pm$ is the total break-up cross section for photons with helicity $\pm 1$.
Again, the requirements for luminosity and control of systematic effects are very stringent, and to date no experiment has determined $A^\gamma_L$. 
However, the continuing developments of high-intensity photon sources put the possibility of measuring $A^\gamma_L$ within reach. In fact, the feasibility of such a measurement is currently being explored for a possible upgrade of the High Intensity Gamma-Ray Source (HI$\gamma$S) at the Triangle Universities Nuclear Laboratory \cite{Ahmed:2013jma}. 

Theoretically, $A^\gamma_L$ (or equivalently $P_\gamma$) has been considered using meson-exchange models \cite{Oka:1983sp,Khriplovich:2000mb,Hyun:2004xp,Liu:2004zm,Schiavilla:2004wn}, effective field theories (EFTs) \cite{Savage:2000iv,Schindler:2009wd,Shin:2009hi}, as well as hybrid methods \cite{Liu:2006dm}. $P_\gamma$ was also considered in Ref.~\cite{Danilov:1965}, which introduced the approach to hadronic parity violation based on the so-called Danilov amplitudes. 
The meson-exchange model results are based on the Desplanques, Donoghue, and Holstein (DDH) framework \cite{Desplanques:1979hn} in combination with various PC models. The DDH approach describes the PV NN interactions in terms of single-meson exchanges of $\pi$, $\rho$, and $\omega$ mesons giving seven phenomenological weak couplings of these mesons to a nucleon. 
It also provides ``best guesses'' and ``reasonable ranges'' based on quark model and symmetry arguments. 
Using the DDH model combined with the Argonne $v_{18}$ (AV18) model for the PC interactions, Ref.~\cite{Liu:2004zm} finds at the photon energy $\omega = 2.235\,\MeV$
\begin{equation}
A^\gamma_L= 2.53 \times 10^{-8}\, .
\end{equation}
This agrees with the result of Ref.~\cite{Schiavilla:2004wn} using the same inputs. 
However, the authors of Ref.~\cite{Schiavilla:2004wn} also show that the result is very sensitive to the values of the PV couplings as well as the choice of PC potential. 
In particular, the asymmetry is larger by a factor of almost two if the CD-Bonn potential \cite{Machleidt:2000ge} is used instead of AV18 (using the same PV parameters).{\footnote{In this comparison the PC couplings in the PV potential were not adjusted to match those used in the PC potential. For a discussion of the impact on the extraction of PV coupling values from experiment see Ref.~\cite{Haxton:2013aca}.} 
Similarly, Ref.~\cite{Hyun:2004xp} shows that the circular polarization $P_\gamma$ depends strongly on the choice of  PC potential.

The asymmetry $A^\gamma_L$ has been determined in pionless EFT (\EFT) at leading order (LO) at threshold in Refs.~\cite{Savage:2000iv,Schindler:2009wd,Shin:2009hi}. 
The philosophy behind parity violation in \EFT is closely related to the approach by Danilov \cite{Danilov:1965} in that no assumptions are made about the short-distance details underlying the mechanism for PV and PC interactions.  Since no model-independent determination of the PV low-energy constants (LECs) of \EFT exists, the authors of \cite{Savage:2000iv,Schindler:2009wd,Shin:2009hi} chose not to give any numerical results.

Given the model dependence of the existing results, the aim of this paper is to analyze the asymmetry $A^\gamma_L$ systematically in \EFT to next-to-leading order (NLO). 
Using \EFT allows one to consistently treat PC and PV interactions in the same framework as well as to provide theoretical error estimates based on its power counting. 
Going beyond the work of Refs.~\cite{Savage:2000iv,Schindler:2009wd}, we determine $A^\gamma_L$ at NLO and study its energy dependence. 
Employing two different conventions for the determination of the PC LECs provides a lower bound on theoretical errors from higher-order effects in the EFT expansion. 
In addition, we consider two very rough ``figures of merit'' to constrain the energy at which an actual measurement of  $A^\gamma_L$ might be best performed. 
As explained below, we use several sets of values of the PV LECs to find an estimate of $A^\gamma_L$. However, these only amount to order-of-magnitude estimates. 
A more detailed and reliable estimate requires the extraction of the LECs from other PV observables, which is not currently feasible. 
Our results show that a measurement of $A^\gamma_L$ should probably be performed for a photon energy $\lesssim 2.3\,\MeV$.
Matching the LECs to the DDH best values we find $A^\gamma_L$ of the order of $10^{-7}$. 
However, given the uncertainty in the PV LECs,  this number should only be considered an order-of-magnitude estimate. 

The paper is structured as follows. In Sec.~\ref{sec:two-body} we introduce the \EFT Lagrangian for the PC and PV sectors. Section~\ref{sec:ampl} contains the definition and calculation of the required PC and PV amplitudes, while results for the asymmetry are found in Sec.~\ref{sec:results}. We conclude in Sec.~\ref{sec:concl}.

\section{Effective Lagrangian}\label{sec:two-body}

Pionless EFT is the effective theory describing interactions between nucleons as well as their couplings to external currents at low momenta well below the pion mass.
The corresponding Lagrangian consists of nucleon contact terms with an increasing number of derivatives. For reviews, see e.g.~Refs.~\cite{vanKolck:1999mw,Beane:2000fx,Platter:2009gz}. In the following, we use the formulation including two auxiliary dibaryon fields with the quantum numbers of two nucleons in the \oneS and \threeS states, respectively \cite{Kaplan:1996nv,Bedaque:1997qi,Beane:2000fi}.
The dibaryon formulation will be denoted by d\EFT. 

The parity-conserving Lagrangian up to NLO in d\EFT is given by
\begin{align}
\mathcal{L}_{PC}^{d}=&\hat{N}^{\dagger}\left(iD_{0}+\frac{\vec{D}^2}{2M_{N}}\right)\hat{N}+\hat{t}_{i}^{\dagger}\left(\Delta^{({}^3\!S_{1})}_{(-1)}+\Delta^{({}^3\!S_{1})}_{(0)}-c_{0t}\left(iD_{0}+\frac{\vec{D}^{2}}{4M_{N}}\right)\right)\hat{t}_{i}-y_{t}\left[\hat{t}_{i}^{\dagger}\hat{N} ^{T}P_{i}\hat{N} +\mathrm{H.c.}\right]\\\nonumber
&+\hat{s}_{a}^{\dagger}\left(\Delta^{({}^1\!S_{0})}_{(-1)}+\Delta^{({}^1\!S_{0})}_{(0)}-c_{0s}\left(iD_{0}+\frac{\vec{D}^{2}}{4M_{N}}\right)\right)\hat{s}_{a}-y_{s}\left[\hat{s}_{a}^{\dagger}\hat{N}^{T}\bar{P}_{a}\hat{N}+\mathrm{H.c.}\right],
\end{align}

\noindent where the dibaryon field $\hat{t}_{i}$ ($\hat{s}_{a}$) is a spin-triplet iso-singlet (spin-singlet iso-triplet) combination of nucleons projected by $P_{i}=\frac{1}{\sqrt{8}}\sigma_{2}\sigma_{i}\tau_{2}$ ($\bar{P}_{a}=\frac{1}{\sqrt{8}}\sigma_{2}\tau_{2}\tau_{a}$). 
The nucleon covariant derivative is
\begin{equation}
D_\mu \hat{N} = \partial_\mu \hat{N} +i \frac{e}{2}(1+\tau_3)\hat{A}_\mu \hat{N} \, ,
\end{equation}
while for the dibaryon fields
\begin{equation}
D_\mu \hat{t}_{i} = \partial_\mu \hat{t}_{i} +i e \hat{A}_\mu \hat{t}_{i} \, , \quad D_\mu \hat{s}_{a} = \partial_\mu \hat{s}_{a} +i e \hat{A}_\mu \mathbf{Q}^{b}_{a}\hat{s}_{b} \, .
\end{equation}
Here $\mathbf{Q}=\mathrm{diag}(2,1,0)$ is a matrix in isospin space acting on the iso-triplet field $\hat{s}$.  The coefficients in the Lagrangian have to be determined by comparison with experimentally accessible quantities, and several conventions exist for this procedure. 
In the so-called $Z$-parametrization \cite{Phillips:1999hh,Griesshammer:2004pe} they are determined by reproducing the \threeS deuteron pole and the \oneS virtual bound state pole at LO, and at NLO one fits to the residues about the \threeS and \oneS poles.  
This scheme yields
\begin{align}
y_{t}&=y_{s}=\sqrt{\frac{4\pi}{M_{N}}}, & \Delta^{(\threeS)}_{(-1)}& =(\gamma_{t}-\mu), & \Delta^{(\oneS)}_{(-1)}&=(\gamma_{s}-\mu), \nonumber \\
 c_{0s/t}&=-\frac{M_{N}}{2\gamma_{s/t}}\left(Z_{s/t}-1\right), & \Delta^{(\threeS)}_{(0)}&=\frac{\gamma_{t}^{2}}{M_{N}},&\Delta^{(\oneS)}_{(0)}&=\frac{\gamma_{s}^{2}}{M_{N}},
\end{align}
where $\gamma_{t}=45.7025$~MeV is the deuteron binding momentum, $\gamma_{s}=-7.890$~MeV the \oneS virtual bound state pole binding momentum, $Z_{t}=1.6908$ the residue about the deuteron pole, and $Z_{s}=.9015$ the residue about the \oneS virtual bound state pole.  
The parameter $\mu$ is  a mass scale given by the power divergence subtraction scheme \cite{Kaplan:1998tg} with dimensional regularization.

The LO \threeS and \oneS dibaryon propagators are given by an infinite bubble sum of nucleons \cite{Kaplan:1998tg}, which at NLO receive corrections from the dibaryon kinetic terms and the $\Delta_{(0)}^{({}^{3}\!S_{1})}$ and $\Delta_{(0)}^{({}^{1}\!S_{0})}$
terms respectively.  The resulting dibaryon propagators in the center of mass (c.m.) frame are \cite{Griesshammer:2004pe}
\begin{equation}
D_{t}\left(\frac{\vect{p}^{2}}{M_{N}},0\right)=\frac{1}{\gamma_{t}+i|\vect{p}|}\left(\underbrace{\vphantom{\frac{Z_{t}-1}{2\gamma_{t}}}1}_{\mathrm{LO}}+\underbrace{\frac{Z_{t}-1}{2\gamma_{t}}(\gamma_{t}-i|\vect{p}|)}_{\mathrm{NLO}}\right),
\end{equation}
and
\begin{equation}
D_{s}\left(\frac{\vect{p}^{2}}{M_{N}},0\right)=\frac{1}{\gamma_{s}+i|\vect{p}|}\left(\underbrace{\vphantom{\frac{Z_{s}-1}{2\gamma_{s}}}1}_{\mathrm{LO}}+\underbrace{\frac{Z_{s}-1}{2\gamma_{s}}(\gamma_{s}-i|\vect{p}|)}_{\mathrm{NLO}}\right),
\end{equation}
where $D_{t}$ is the \threeS dibaryon propagator and $D_{s}$ the \oneS dibaryon propagator.  From the deuteron propagator we obtain the deuteron wavefunction renormalization as the residue about its pole which yields
\begin{equation}
Z_{D}=\frac{8\pi\gamma_{t}}{M_{N}^{2}y_{t}^{2}}\left[\underbrace{\vphantom{Z_{t}-1}1}_{\mathrm{LO}}+\underbrace{Z_{t}-1}_{\mathrm{NLO}}\right].
\end{equation}

In addition to the $Z$-parametrization we also consider another formalism, which we refer to as the resummed effective range expansion (ERE).  
In this approach effective range corrections are treated as LO terms and are resummed into the LO dibaryon propagators yielding
\begin{equation}
D_{t}\left(\frac{\vect{p}^{2}}{M_{N}},0\right)=\frac{1}{\gamma_{t}-\frac{1}{2}\rho_{t}(\vect{p}^{2}+\gamma_{t}^2)+i|\vect{p}|},
\end{equation}
and
\begin{equation}
D_{s}\left(\frac{\vect{p}^{2}}{M_{N}},0\right)=\frac{1}{\gamma_{s}-\frac{1}{2}r_{s}\vect{p}^{2}+i|\vect{p}|},
\end{equation}
where $\rho_{t}=1.764$~fm is the effective range about the deuteron pole and $r_{s}=2.73$~fm is the effective range about zero momentum in the \oneS channel. 
Note that in the resummed ERE, $\gamma_{s}=\frac{1}{a_{s}}$, where $a_{s}=-23.714$~fm is the \oneS scattering length and the deuteron wavefunction renormalization is given by 
\begin{equation}
Z_{D}=\frac{8\pi\gamma_{t}}{M_{N}^{2}y_{t}^{2}}Z_{t}.
\end{equation}
The PC LECs are given by
\begin{align}
& \Delta^{(\threeS)}_{(0)}=\frac{\gamma_{t}^{2}}{M_{N}}, \quad \Delta^{(\oneS)}_{(0)}=0, \quad c_{0t}=-\frac{\rho_{t}}{2}, \quad c_{0s}=-\frac{r_{s}}{2},
\end{align}
while the other LECs remain the same as in the $Z$-parametrization.\footnote{This is not the most conventional choice of LECs in the resummed ERE formalism (see e.g.~Ref.~\cite{Beane:2000fi}). However, physical results are independent of which convention is used.}

The nucleon and dibaryon fields can couple to external electromagnetic currents through the covariant derivative. In addition, the nucleon can couple through its magnetic dipole moment. The corresponding LO Lagrangian is given by 
\begin{equation}
\label{eq:LOMagPhoton}
\mathcal{L}_\kappa=\frac{e}{2M_{N}}\hat{N}^{\dagger}(\kappa_{0}+\kappa_{1}\tau_{3})\vec{\sigma}\cdot \mathbf{B}\hat{N},
\end{equation}
where $\kappa_0 = 0.4399$ is the isoscalar magnetic dipole moment, and  $\kappa_1 = 2.3529$ the isovector magnetic dipole moment.  
At NLO there is also a four-nucleon-one-photon contact interaction with a coupling constant $L_{1}$ \cite{Chen:1999tn} which in d\EFT is \cite{Beane:2000fi}
\begin{equation}
\label{eq:NLOMagPhoton}
\mathcal{L}_{L_1}^d=e\frac{L_{1}y_{s}y_{t}M_{N}}{8\pi}\hat{t}^{j\dagger}\hat{s}_{3}\mathbf{B}_{j}+\mathrm{H.c.}
\end{equation}
The constant $L_{1}$ is fit to reproduce the $np$ capture cross section of $\sigma^{expt}=334.2\pm .5$ mb at a neutron velocity of 2200 m/s.  At the same order, there is an additional four-nucleon-one-photon contact interaction proportional to a LEC $L_{2}$ \cite{Chen:1999tn,Kaplan:1998sz}  in d\EFT which is given by the Lagrangian 
\begin{equation}
\label{eq:L2}
\mathcal{L}_{L_2}^d=-e\frac{L_{2}y_{t}^{2}M_{N}}{8\pi}i\epsilon^{ijk}\hat{t}_{i}^{\dagger}\hat{t}_{j}\mathbf{B}_{k}.
\end{equation}
The value for $L_{2}$ is fit to reproduce the correct deuteron magnetic dipole moment at NLO \cite{Chen:1999vd}.

The LO PV Lagrangian in \EFT consists of five independent terms \cite{Girlanda:2008ts,Phillips:2008hn} and in the dibaryon formalism is given by \cite{Schindler:2009wd}
\begin{align}
\label{eq:PVLag}
\mathcal{L}^{d}_{PV}=-&\left[g^{({}^{3}\!S_{1}-{}^{1}\!P_{1})}(\hat{t}_{i})^{\dagger}\left(\hat{N}^{T}\sigma_{2}\tau_{2}i\LRd_{i}\hat{N}\right)\right.\\\nonumber
&+g^{({}^{1}\!S_{0}-{}^{3}\!P_{0})}_{(\Delta I=0)}(\hat{s}_{a})^{\dagger}\left(\hat{N}^{T}\sigma_{2}\vec{\sigma}\cdot\tau_{2}\tau_{a}i\LRd\hat{N}\right)\\\nonumber
&+g^{({}^{1}\!S_{0}-{}^{3}\!P_{0})}_{(\Delta I=1)}\epsilon^{3ab}(\hat{s}^{a})^{\dagger}\left(\hat{N}^{T}\sigma_{2}\vec{\sigma}\cdot\tau_{2}\tau^{b}\LRd\hat{N}\right)\\\nonumber
&+g^{({}^{1}\!S_{0}-{}^{3}\!P_{0})}_{(\Delta I=2)}\mathcal{I}^{ab}(\hat{s}^{a})^{\dagger}\left(\hat{N}^{T}\sigma_{2}\vec{\sigma}\cdot\tau_{2}\tau^{b}i\LRd\hat{N}\right)\\\nonumber
&+\left.g^{({}^{3}\!S_{1}-{}^{3}\!P_{1})}\epsilon^{ijk}(\hat{t}_{i})^{\dagger}\left(\hat{N}^{T}\sigma_{2}\sigma^{k}\tau_{2}\tau_{3}\LRd^{j}\hat{N}\right)\right]+\mathrm{H.c.}
\end{align}
For the isotensor contribution of the PV Lagrangian the matrix $\mathcal{I}_{ab}$ is given by $\mathcal{I}_{ab}=\mathrm{diag}(1,1,-2)$.  The notation $\LRd_{i}$ is defined as $a\, \mathcal{O}\LRd b = a\,\mathcal{O}\vec D b - (\vec D a)\mathcal{O} b$, where $\mathcal{O}$ is some spin-isospin operator.  Higher-order PV operators contain at least two more derivatives and thus only start to enter at next-to-next-to-leading order ($\mathrm{N}^{2}\mathrm{LO}$).


\section{Amplitudes}\label{sec:ampl}

The amplitude for $np\to d\gamma$ including PC and PV terms can be parametrized as
\begin{align}
\mathcal{A}&=eXN^{T}\tau_{2}\sigma_{2}[\vect{\sigma}\cdot\vect{k}\pDeut\cdot\pGamma-\vect{\sigma}\cdot\pGamma\vect{k}\cdot\pDeut]N +ieY\epsilon^{ijk}\pDeutm{i}\vect{k}_{j}\pGammam{k}(N^{T}\tau_{2}\tau_{3}\sigma_{2}N)\\\nonumber
&+eE1_{v}N^{T}\sigma_{2}\vect{\sigma}\cdot\pDeut\tau_{2}\tau_{3} N\vect{p}\cdot\pGamma +ieW\epsilon^{ijk}\pDeutm{i}\pGammam{k}(N^{T}\tau_{2}\sigma_{2}\sigma^{j}N)+eV\pDeut\cdot\pGamma(N^{T}\tau_{2}\tau_{3}\sigma_{2}N)\\\nonumber
&+ieU_{1}\epsilon^{ijk}\pGammam{i}\,k_{j}\pDeutm{k}N^{T}\sigma_{2}\vect{\sigma}\cdot \vect{p}\tau_{2}\tau_{3}N+ieU_{2}\epsilon^{ijk}\pGammam{i}\,k_{j}N^{T}\sigma_{2}\sigma_{k}\tau_{2}\tau_{3}N\vect{p}\cdot\pDeut\\\nonumber
&+ieU_{3}\epsilon^{ijk}\pGammam{i}\,k_{j}N^{T}\sigma_{2}\boldsymbol{\sigma}\cdot\pDeut\tau_{2}\tau_{3}Np_{k}+\cdots,
\end{align}
where the ellipsis stands for terms not relevant for our calculation.  Here $\vect{k}$ represents the outgoing photon momentum and $\vect{p}$ the nucleon momentum in the center of mass frame.  The polarization of the deuteron and photon are defined by $\hat{\epsilon}_{d}$ and $\hat{\epsilon}_{\gamma}$ respectively, and $N$ defines the nucleon spinor and isospinor.  
We use the convention that $\hat\epsilon_\gamma^\pm=\mp(1,\pm i,0)/\sqrt{2}$ is the polarization vector for photons with $\pm$ helicity.  $X$ denotes the isoscalar magnetic dipole (M1) amplitude, $Y$ the isovector M1 amplitude, and $E1_{v}$ the isovector electric dipole (E1) amplitude.  
The PV amplitudes are the PV isoscalar E1 amplitude $W$, the PV isovector E1 amplitude $V$, and three PV isovector M1 amplitudes $U_{1}$, $U_{2}$, and $U_{3}$.\footnote{These amplitudes are also sometimes referred to as multipole moments.}
In \EFT, each amplitude can be decomposed into contributions at a given order, e.g. $Y=Y_{\LO}+Y_{\NLO}+\cdots$, with analogous expressions for the remaining amplitudes.  The PC amplitudes have been calculated in various formalisms and conventions elsewhere.
Here, we collect the results in our conventions before considering the energy dependence of the PV amplitudes $V$, $W$, $U_{1}$, $U_{2}$, and $U_{3}$ up to NLO.

\subsection{Parity-conserving amplitudes}

The isovector electric dipole amplitude $E1_v$ in d\EFT at LO is given by the diagram in Fig.~\ref{fig:E1v}, in which photons are minimally coupled to the nucleons.
\begin{figure}[hbt]
\includegraphics[width=40mm]{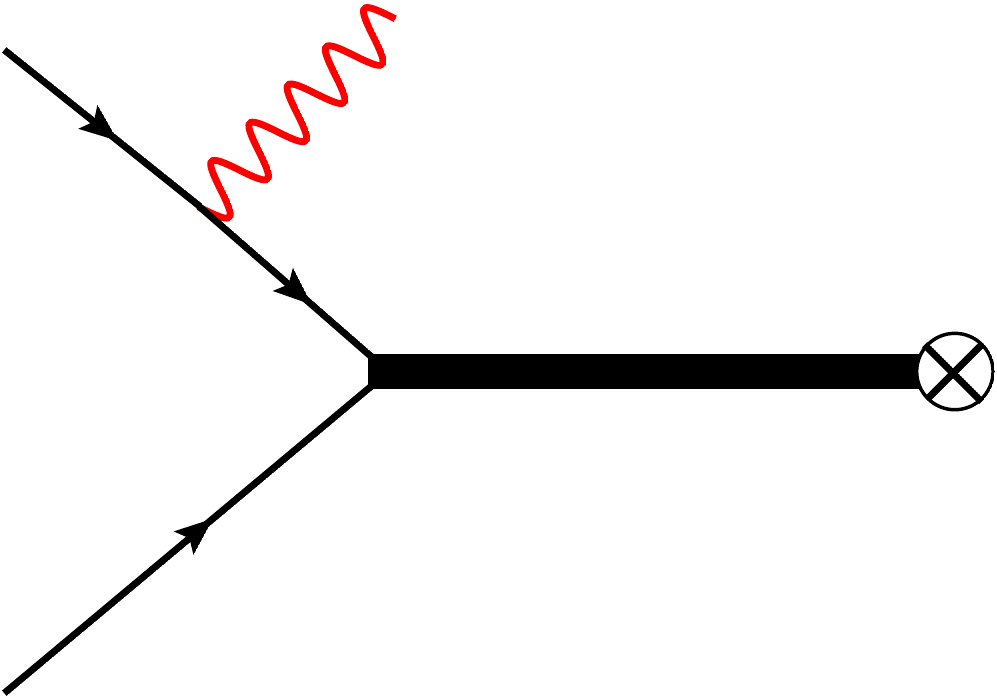}
\caption{\label{fig:E1v}(Color online) Diagram contributing to the $E1_{v}$ amplitude.  The thick solid line denotes a dibaryon propagator, the thin lines with arrows nucleon lines, and the wavy line a photon.  The photon is minimally coupled and the cross denotes an insertion of the deuteron wavefunction renormalization.}
\end{figure}
At NLO there are no new diagrams for $E1_{v}$ and corrections in the $Z$-parametrization only come from the deuteron wavefunction renormalization.
Using the ERE parametrization, the $E1_{v}$ amplitude has been calculated in \EFT up to $\mathrm{N}^{4}\mathrm{LO}$ \cite{Rupak:1999rk,Chen:1999bg}, and in d\EFT in the resummed ERE up to NLO \cite{Ando:2005cz}.  
The LO $E1_{v}$ amplitude in the $Z$-parametrization is given by
\begin{equation}
{E1_{v,\LO}^{(Z)}}=-\frac{2}{M_{N}}\sqrt{\gamma_{t}\pi}\frac{1}{\gamma_{t}^{2}+\vect{p}^{2}},
\end{equation}
and in the resummed ERE by
\begin{equation}
{E1_{v,\LO}^{(R)}}=-\frac{2}{M_{N}}\sqrt{\gamma_{t}\pi}\frac{1}{\gamma_{t}^{2}+\vect{p}^{2}}\sqrt{Z_{t}}.
\end{equation}
The strictly perturbative NLO correction in the $Z$-parametrization is\
\begin{equation}
{E1_{v}}^{(Z)}_{\NLO}=-\frac{2}{M_{N}}\sqrt{\gamma_{t}\pi}\frac{1}{\gamma_{t}^{2}+\vect{p}^{2}}\frac{1}{2}(Z_{t}-1),
\end{equation}
and in the resummed ERE the NLO term is
\begin{equation}
{E1_{v}}^{(R)}_{\NLO}=0.
\end{equation}

At LO and NLO the isovector M1 amplitude $Y$ is given by the sum of diagrams in Fig.~\ref{fig:Y}.
\begin{figure}[hbt]
\includegraphics[width=100mm]{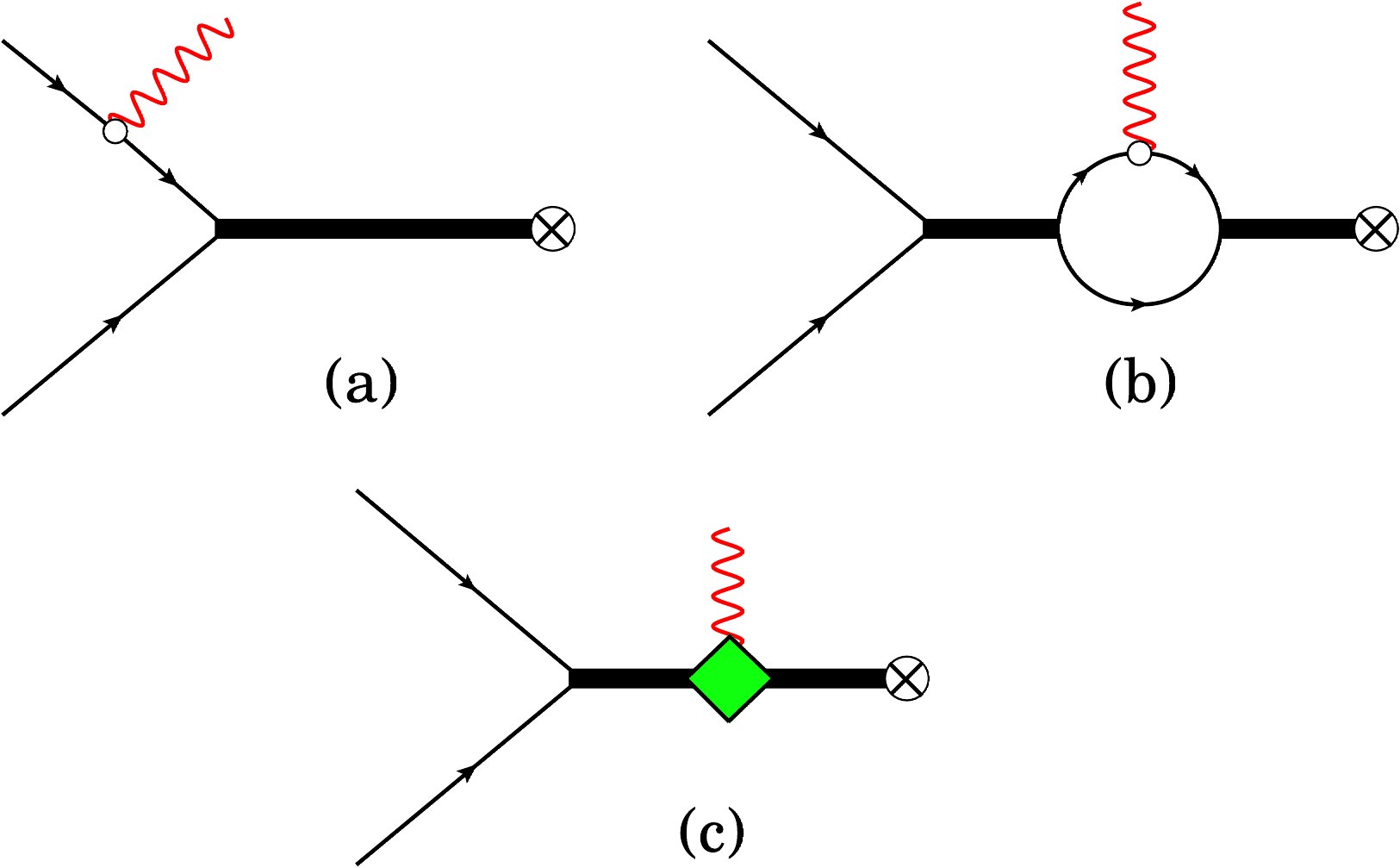}
\caption{\label{fig:Y}(Color online) Diagrams contributing to the amplitude $Y$.  Here the small open circle denotes a coupling to the nucleon magnetic moment.  The green diamond represents an insertion of $L_{1}$.}
\end{figure}
The LO contributions are given by diagrams (a) and (b) where the photon couples magnetically via the Lagrangian of Eq.~(\ref{eq:LOMagPhoton}). 
In the NLO diagram (c) the photon couples through the $L_{1}$ term from the Lagrangian of Eq.~(\ref{eq:NLOMagPhoton}).  
In addition to this NLO diagram, in the $Z$-parametrization, the LO diagrams (a) and (b) receive NLO corrections from the deuteron wavefunction renormalization as well as effective range corrections to the dibaryon propagators.  
Using the ERE parametrization the amplitude $Y$ has been calculated to $\mathrm{N}^{2}$LO in \EFT \cite{Rupak:1999rk,Chen:1999tn} and to NLO in d\EFT in the resummed ERE \cite{Ando:2004mm,Beane:2000fi}.  
Summing the diagrams we find in the $Z$-parametrization a LO $Y$ amplitude
\begin{align}
Y^{(Z)}_{\LO}=&\frac{2\kappa_{1}}{M_{N}}\sqrt{\gamma_{t}\pi}\frac{1}{\gamma_{t}^{2}+\vect{p}^{2}}\left(1-\frac{\gamma_{t}+i|\vect{p}|}{\gamma_{s}+i|\vect{p}|}\right),
\end{align}
and in the resummed ERE
\begin{align}
Y^{(R)}_{\LO}=&\frac{2\kappa_{1}}{M_{N}}\sqrt{\gamma_{t}\pi}\frac{1}{\gamma_{t}^{2}+\vect{p}^{2}}\left(1-\frac{\gamma_{t}+i|\vect{p}|}{\gamma_{s}-\frac{1}{2}r_{s}\vect{p}^{2}+i|\vect{p}|}\right)\sqrt{Z_{t}} \, .
\end{align}
The NLO $Z$-parametrization correction is
\begin{align}
Y^{(Z)}_{\NLO}=&\frac{2\kappa_{1}}{M_{N}}\sqrt{\gamma_{t}\pi}\frac{1}{\gamma_{t}^{2}+\vect{p}^{2}}\left(\frac{1}{2}(Z_{t}-1)\right.\\\nonumber
&\left.-\frac{1}{\gamma_{s}+i|\vect{p}|}\left\{\frac{1}{2}(Z_{t}-1)+\frac{Z_{s}-1}{2\gamma_{s}}(\gamma_{s}-i|\vect{p}|)\right\}(\gamma_{t}+i|\vect{p}|)\right) \\\nonumber
&-\frac{L_{1}}{M_{N}}\sqrt{\gamma_{t}\pi}\frac{1}{\gamma_{s}+i|\vect{p}|},
\end{align}
and the NLO resummed ERE correction is
\begin{align}
Y^{(R)}_{\NLO}=-\frac{L_{1}}{M_{N}}\sqrt{\gamma_{t}\pi}\frac{1}{\gamma_{s}-\frac{1}{2}r_{s}\vect{p}^{2}+i|\vect{p}|}\sqrt{Z_{t}} \, .
\end{align}

In \EFT, there are no LO contributions to the isoscalar M1 amplitude $X$ in the zero recoil limit \cite{Savage:1998ae}. The first nonzero contribution occurs at NLO from the four-nucleon-one-photon LEC $L_{2}$ \cite{Chen:1999vd} given in Eq.~(\ref{eq:L2}).  The NLO contribution to $X$ is given by the sole diagram in Fig.~\ref{fig:X}, where the circle represents an $L_{2}$ vertex.
\begin{figure}[hbt]
\includegraphics[width=40mm]{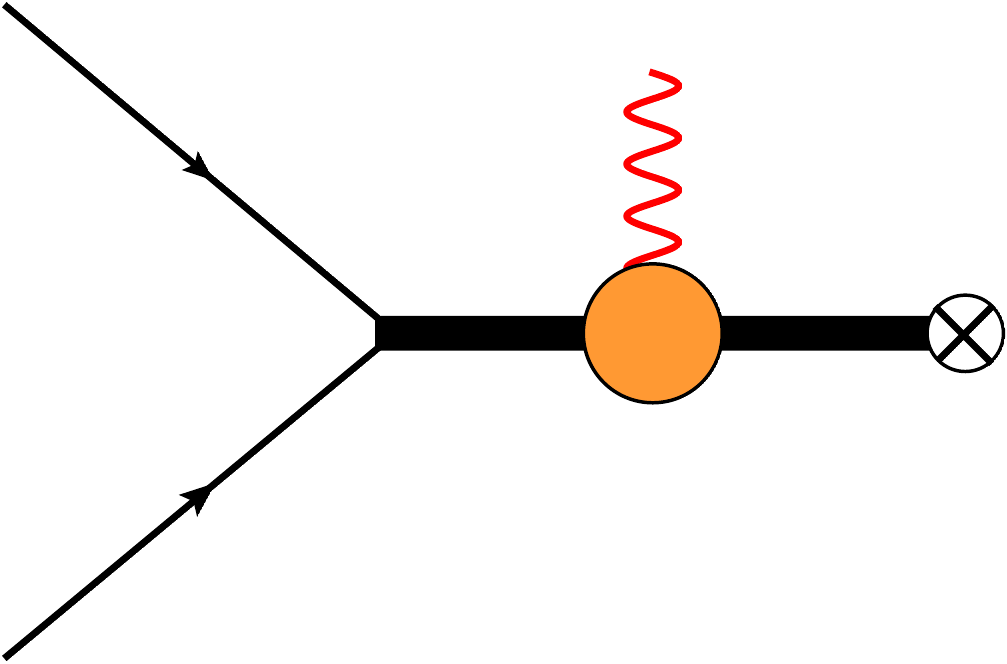}
\caption{\label{fig:X}(Color online) NLO diagram contributing to the amplitude $X$.  The orange circle represents an insertion of $L_{2}$.}
\end{figure}
The resulting $X$ at NLO in d\EFT in the $Z$-parametrization is given by
\begin{equation}
X^{(Z)}_{\NLO}=\frac{L_{2}}{M_{N}}\sqrt{\gamma_{t}\pi}\frac{1}{\gamma_{t}+i|\vect{p}|},
\end{equation}
and in the resummed ERE by
\begin{equation}
X^{(R)}_{\NLO}=\frac{L_{2}}{M_{N}}\sqrt{\gamma_{t}\pi}\frac{1}{\gamma_{t}-\frac{1}{2}\rho_{t}(\vect{p}^{2}+\gamma_{t}^{2})+i|\vect{p}|}\sqrt{Z_{t}} \, .
\end{equation}

\subsection{Parity-violating amplitudes}

The PV amplitudes V and W are given by the sum of diagrams in Fig.~\ref{fig:PV}, where the square vertex represents an insertion from the PV Lagrangian of Eq.~\eqref{eq:PVLag}.  The initial NN state and the first dibaryon propagators in diagrams
\ref{fig:PV}(a)-(d) are  in the \threeS-wave for the amplitude $W$ and in the \oneS-wave for $V$. 
Diagrams (a)-(e) have been calculated in the resummed ERE parametrization in \EFT and d\EFT at LO in the threshold limit in Refs.~\cite{Savage:2000iv,Schindler:2009wd,Shin:2009hi}.  
Here we calculate these amplitudes beyond threshold in the $Z$-parametrization and resummed ERE up to NLO. Diagram (f) is zero in the threshold limit and thus was not considered in previous calculations.
\begin{figure}[hbt]
\includegraphics[width=100mm]{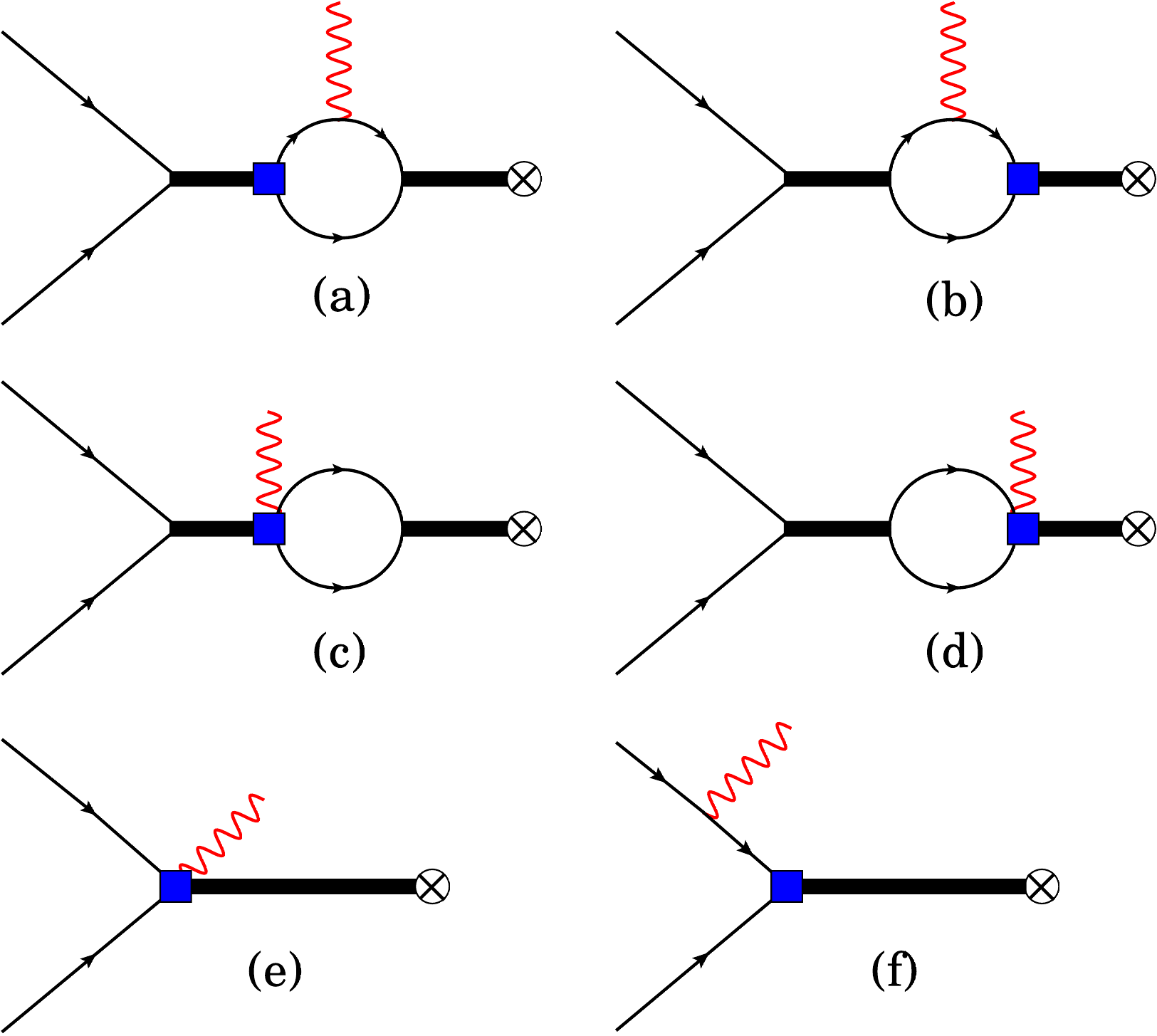}
\caption{\label{fig:PV}(Color online) LO diagrams contributing to the PV amplitudes $W$ and $V$.  The blue box represents an insertion of a PV operator.}
\end{figure}
For the $V$ and $W$ amplitudes no new types of diagrams enter at NLO.
In the $Z$-parametrization the NLO contributions stem from effective range corrections to the dibaryon propagators and the deuteron wavefunction renormalization.  For the LO PV isoscalar E1 amplitude $W$ in the $Z$-parametrization we find
\begin{align}
\label{eq:WLO}
&W^{(Z)}_{\LO}=-\sqrt{\frac{8\gamma_{t}}{M_{N}}}\left(1-\frac{1}{3}\frac{1}{\gamma_{t}^{2}+\vect{p}^{2}}\left\{\vect{p}^{2}+\frac{(\gamma_{t}+i|\vect{p}|)^{3}}{\gamma_{t}+i\vect{p}}\right\}\right)g^{{}^{3}\!S_{1}-{}^{3}\!P_{1}},
\end{align}
and in the resummed ERE
\begin{align}
&W^{(R)}_{\LO}=-\sqrt{\frac{8\gamma_{t}Z_{t}}{M_{N}}}\left(1-\frac{1}{3}\frac{1}{\gamma_{t}^{2}+\vect{p}^{2}}\left\{\vect{p}^{2}+\frac{(\gamma_{t}+i|\vect{p}|)^{3}}{\gamma_{t}-\frac{1}{2}\rho_{t}(\vect{p}^{2}+\gamma_{t}^{2})+i\vect{p}}\right\}\right)g^{{}^{3}\!S_{1}-{}^{3}\!P_{1}}.
\end{align}
The NLO $Z$-parametrization correction is
\begin{align}
\label{eq:WNLO}
&W^{(Z)}_{\NLO}=-\sqrt{\frac{8\gamma_{t}}{M_{N}}}\left(1-\frac{1}{3}\frac{1}{\gamma_{t}^{2}+\vect{p}^{2}}\left\{\vect{p}^{2}+\frac{(\gamma_{t}+i|\vect{p}|)^{3}}{\gamma_{t}+i\vect{p}}\left(1+\frac{1}{\gamma_{t}}(\gamma_{t}-i|\vect{p}|)\right)\right\}\right)\frac{Z_{t}-1}{2}g^{{}^{3}\!S_{1}-{}^{3}\!P_{1}},
\end{align}
and the NLO resummed ERE correction is
\begin{align}
&W^{(R)}_{\NLO}=0.
\end{align}
The LO PV isovector E1 amplitude $V$ in the $Z$-parametrization is 
\begin{equation}
\label{eq:VLO}
\begin{split}
V^{(Z)}_{\LO}=&-\sqrt{\frac{8\gamma_{t}}{M_{N}}}\left[\left(1-\frac{1}{\gamma_{s}+i|\vect{p}|}\left\{i|\vect{p}|+\frac{2}{3}\frac{\gamma_{t}^{3}-i|\vect{p}|^{3}}{\gamma_{t}^{2}+\vect{p}^{2}}\right\}-\frac{1}{3}\frac{\vect{p}^{2}}{\gamma_{t}^{2}+\vect{p}^{2}}\right)g^{{}^{3}\!S_{1}-{}^{1}\!P_{1}}\right.\\
&\left.+\frac{1}{\gamma_{s}+i|\vect{p}|}\left\{\gamma_{t}-\frac{2}{3}\frac{\gamma_{t}^{3}-i|\vect{p}|^{3}}{\gamma_{t}^{2}+\vect{p}^{2}}\right\}\left(g^{{}^{1}\!S_{0}-{}^{3}\!P_{0}}_{(\Delta I=0)}-2g^{{}^{1}\!S_{0}-{}^{3}\!P_{0}}_{(\Delta I=2)}\right)\right],
\end{split}
\end{equation}
and in the resummed ERE
\begin{equation}
\label{eq:VLOERE}
\begin{split}
V^{(R)}_{\LO}=&-\sqrt{\frac{8\gamma_{t}Z_{t}}{M_{N}}}\left[\left(1-\frac{1}{\gamma_{s}-\frac{1}{2}r_{s}\vect{p}^{2}+i|\vect{p}|}\left\{i|\vect{p}|+\frac{2}{3}\frac{\gamma_{t}^{3}-i|\vect{p}|^{3}}{\gamma_{t}^{2}+\vect{p}^{2}}\right\}-\frac{1}{3}\frac{\vect{p}^{2}}{\gamma_{t}^{2}+\vect{p}^{2}}\right)g^{{}^{3}\!S_{1}-{}^{1}\!P_{1}}\right.\\
&\left.+\frac{1}{\gamma_{s}-\frac{1}{2}r_{s}\vect{p}^{2}+i|\vect{p}|}\left\{\gamma_{t}-\frac{2}{3}\frac{\gamma_{t}^{3}-i|\vect{p}|^{3}}{\gamma_{t}^{2}+\vect{p}^{2}}\right\}\left(g^{{}^{1}\!S_{0}-{}^{3}\!P_{0}}_{(\Delta I=0)}-2g^{{}^{1}\!S_{0}-{}^{3}\!P_{0}}_{(\Delta I=2)}\right)\right].
\end{split}
\end{equation}
The NLO $Z$-parametrization correction is
\begin{equation}
\label{eq:VNLO}
\begin{split}
&V^{(Z)}_{\NLO}= -\sqrt{\frac{8\gamma_{t}}{M_{N}}}\left[\left(\frac{1}{2}(Z_{t}-1)\right.\right.\\
&\left.-\frac{1}{\gamma_{s}+i|\vect{p}|}\left(\frac{1}{2}(Z_{t}-1)+\frac{Z_{s}-1}{2\gamma_{s}}(\gamma_{s}-i|\vect{p}|)\right)\left\{i|\vect{p}|+\frac{2}{3}\frac{\gamma_{t}^{3}-i|\vect{p}|^{3}}{\gamma_{t}^{2}+\vect{p}^{2}}\right\}-\frac{1}{3}\frac{\vect{p}^{2}}{\gamma_{t}^{2}+\vect{p}^{2}}\frac{1}{2}(Z_{t}-1)\right)g^{{}^{3}\!S_{1}-{}^{1}\!P_{1}}\\
&\left.+\frac{1}{\gamma_{s}+i|\vect{p}|}\left(\frac{1}{2}(Z_{t}-1)+\frac{Z_{s}-1}{2\gamma_{s}}(\gamma_{s}-i|\vect{p}|)\right)\left\{\gamma_{t}-\frac{2}{3}\frac{\gamma_{t}^{3}-i|\vect{p}|^{3}}{\gamma_{t}^{2}+\vect{p}^{2}}\right\}\left(g^{{}^{1}\!S_{0}-{}^{3}\!P_{0}}_{(\Delta I=0)}-2g^{{}^{1}\!S_{0}-{}^{3}\!P_{0}}_{(\Delta I=2)}\right)\right],
\end{split}
\end{equation}
and in the resummed ERE
\begin{equation}
V^{(R)}_{\NLO}=0.
\end{equation}

The LO PV amplitudes, $U_{1}$, $U_{2}$, and $U_{3}$ receive contributions from the diagrams in Fig \ref{fig:PVMore}.  In the zero recoil limit at threshold these diagrams give zero contribution and therefore did not appear in previous calculations.
\begin{figure}[hbt]
\includegraphics[width=100mm]{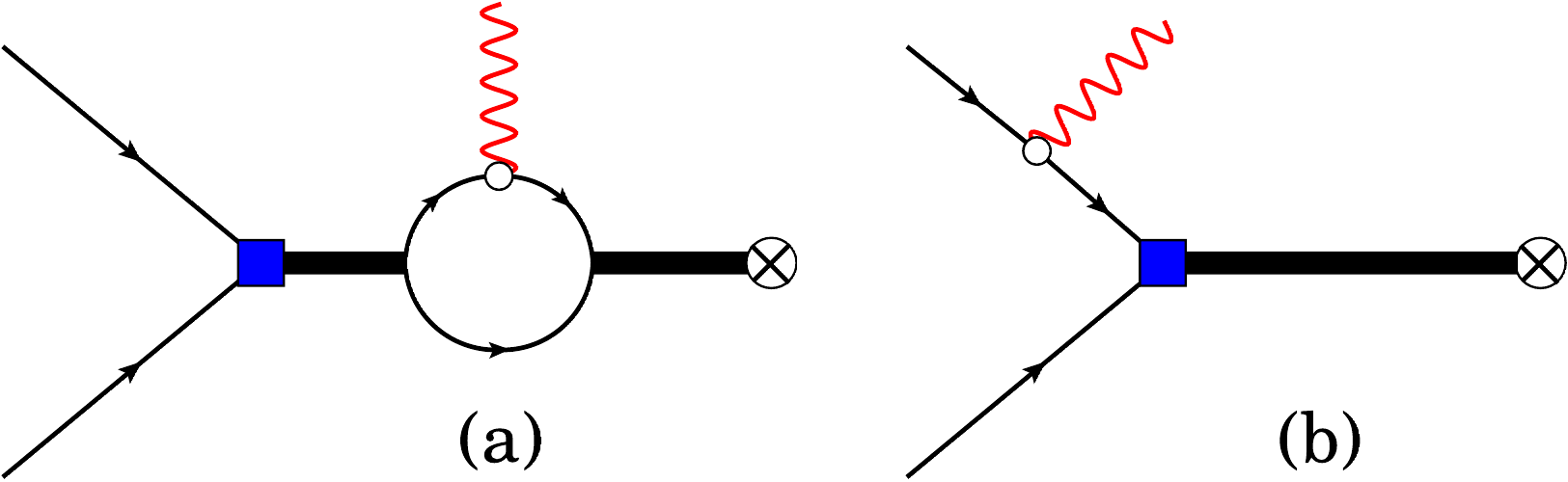}
\caption{\label{fig:PVMore}(Color online) LO diagrams contributing to the PV amplitudes $U_1$, $U_2$, and $U_3$. The blue box represents an insertion of a PV operator.}
\end{figure}
The LO PV isovector M1 amplitude $U_{1}$ in the $Z$-parametrization is
\begin{equation}
{U_{1}}^{(Z)}_{LO}=\sqrt{\frac{8\gamma_{t}}{M_{N}}}\frac{1}{\gamma_{t}^{2}+\vect{p}^{2}}\left(\kappa_{0}g^{{}^{3}\!S_{1}-{}^{3}\!P_{1}}-2\kappa_{1}\frac{\gamma_{t}+i\vect{p}}{\gamma_{s}+i\vect{p}}\left(g^{{}^{1}\!S_{0}-{}^{3}\!P_{0}}_{(\Delta I=0)}-2g^{{}^{1}\!S_{0}-{}^{3}\!P_{0}}_{(\Delta I=2)}\right)\right),
\end{equation}
and in the resummed ERE
\begin{equation}
{U_{1}}^{(R)}_{LO}=\sqrt{\frac{8\gamma_{t}Z_{t}}{M_{N}}}\frac{1}{\gamma_{t}^{2}+\vect{p}^{2}}\left(\kappa_{0}g^{{}^{3}\!S_{1}-{}^{3}\!P_{1}}-2\kappa_{1}\frac{\gamma_{t}+i\vect{p}}{\gamma_{s}-\frac{1}{2}r_{s}\vect{p}^{2}+i\vect{p}}\left(g^{{}^{1}\!S_{0}-{}^{3}\!P_{0}}_{(\Delta I=0)}-2g^{{}^{1}\!S_{0}-{}^{3}\!P_{0}}_{(\Delta I=2)}\right)\right).
\end{equation}
The LO PV isovector M1 amplitude $U_{2}$ in the $Z$-parametrization is
\begin{equation}
{U_{2}}^{(Z)}_{LO}=\sqrt{\frac{8\gamma_{t}}{M_{N}}}\frac{1}{\gamma_{t}^{2}+\vect{p}^{2}}\left(\kappa_{1}g^{{}^{3}\!S_{1}-{}^{1}\!P_{1}}-2\kappa_{0}g^{{}^{3}\!S_{1}-{}^{3}\!P_{1}}\right),
\end{equation}
and in the resummed in the ERE
\begin{equation}
{U_{2}}^{(R)}_{LO}=\sqrt{\frac{8\gamma_{t}Z_{t}}{M_{N}}}\frac{1}{\gamma_{t}^{2}+\vect{p}^{2}}\left(\kappa_{1}g^{{}^{3}\!S_{1}-{}^{1}\!P_{1}}-2\kappa_{0}\frac{\gamma_{t}+i\vect{p}}{\gamma_{t}-\frac{1}{2}\rho_{t}(\vect{p}^{2}+\gamma_{t}^{2})+i\vect{p}}g^{{}^{3}\!S_{1}-{}^{3}\!P_{1}}\right).
\end{equation}
Finally the amplitude $U_{3}$ in the $Z$-parametrization is given by
\begin{equation}
{U_{3}}^{(Z)}_{LO}=\sqrt{\frac{8\gamma_{t}}{M_{N}}}\frac{1}{\gamma_{t}^{2}+\vect{p}^{2}}\kappa_{0}g^{{}^{3}\!S_{1}-{}^{3}\!P_{1}},
\end{equation}
and in the resummed ERE
\begin{equation}
{U_{3}}^{(R)}_{LO}=\sqrt{\frac{8\gamma_{t}Z_{t}}{M_{N}}}\frac{1}{\gamma_{t}^{2}+\vect{p}^{2}}\left(2\frac{\gamma_{t}+i\vect{p}}{\gamma_{t}-\frac{1}{2}\rho_{t}(\vect{p}^{2}+\gamma_{t}^{2})+i\vect{p}}-1\right)\kappa_{0}g^{{}^{3}\!S_{1}-{}^{3}\!P_{1}}.
\end{equation}

At NLO the amplitudes $U_{1}$, $U_{2}$, and $U_{3}$ receive wavefunction renormalization corrections as well as corrections to the dibaryon propagators.  In addition, the diagrams shown in Fig.~\ref{fig:PVNLO}, resulting from a combination of a PV vertex with the $L_{1}$ and $L_{2}$ interactions, contribute at NLO.
\begin{figure}[hbt]
\includegraphics[width=100mm]{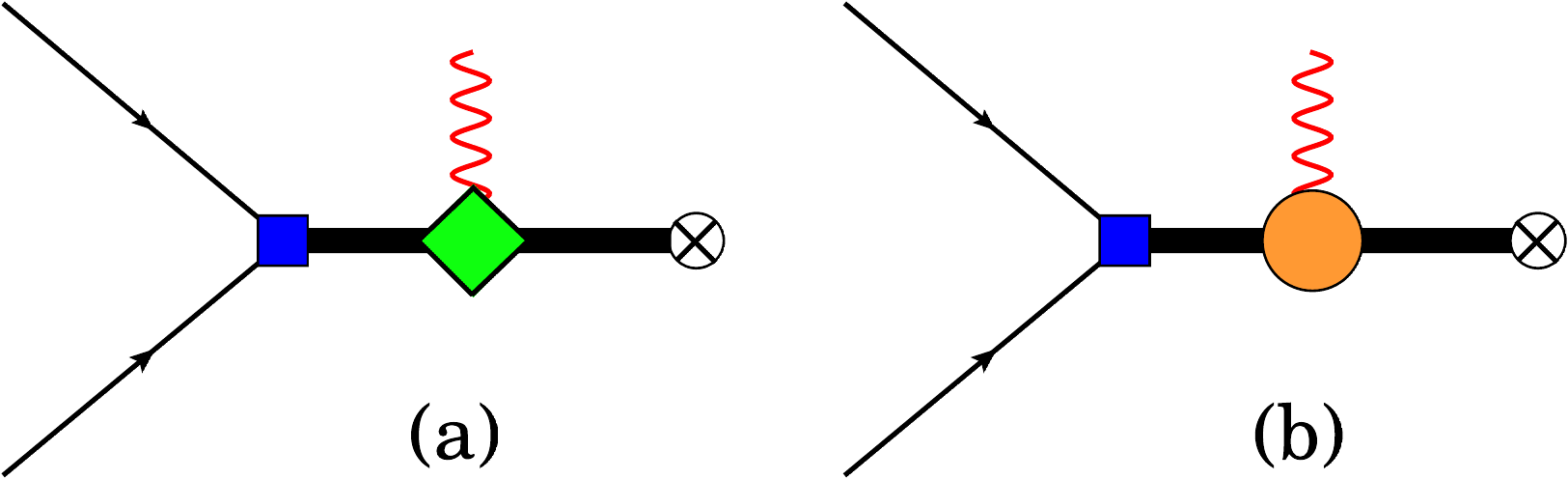}
\caption{\label{fig:PVNLO}(Color online) NLO diagrams contributing to the PV amplitudes $U_{1}$, $U_{2}$, and $U_{3}$.  The blue box represents an insertion of a PV operator.  The green diamond represents an insertion of $L_{1}$ and the orange circle an insertion of $L_{2}$.}
\end{figure}
The NLO $Z$-parametrization contribution to $U_{1}$ is 
\begin{align}
&{U_{1}}^{(Z)}_{NLO}=\sqrt{\frac{8\gamma_{t}}{M_{N}}}\frac{1}{\gamma_{t}^{2}+\vect{p}^{2}}\left(\kappa_{0}g^{{}^{3}\!S_{1}-{}^{3}\!P_{1}}\frac{1}{2}(Z_{t}-1)\right.\\\nonumber
&\left.-2\kappa_{1}\frac{\gamma_{t}+i\vect{p}}{\gamma_{s}+i\vect{p}}\left(\frac{1}{2}(Z_{t}-1)+\frac{Z_{s}-1}{2\gamma_{s}}(\gamma_{s}-i|\vect{p}|)\right)\left(g^{{}^{1}\!S_{0}-{}^{3}\!P_{0}}_{(\Delta I=0)}-2g^{{}^{1}\!S_{0}-{}^{3}\!P_{0}}_{(\Delta I=2)}\right)\right)\\\nonumber
&-L_{1}\sqrt{\frac{8\gamma_{t}}{M_{N}}}\frac{1}{\gamma_{s}+i\vect{p}}\left(g^{{}^{1}\!S_{0}-{}^{3}\!P_{0}}_{(\Delta I=0)}-2g^{{}^{1}\!S_{0}-{}^{3}\!P_{0}}_{(\Delta I=2)}\right),
\end{align}
and in the resummed ERE
\begin{equation}
{U_{1}}_{\NLO}^{(R)}=-L_{1}\sqrt{\frac{8\gamma_{t}Z_{t}}{M_{N}}}\frac{1}{\gamma_{s}-\frac{1}{2}r_{s}p^{2}+i\vect{p}}\left(g^{{}^{1}\!S_{0}-{}^{3}\!P_{0}}_{(\Delta I=0)}-2g^{{}^{1}\!S_{0}-{}^{3}\!P_{0}}_{(\Delta I=2)}\right).
\end{equation}
The NLO $Z$-parametrization contribution to $U_{2}$ is
\begin{align}
{U_{2}}^{(Z)}_{NLO}=&\sqrt{\frac{8\gamma_{t}}{M_{N}}}\frac{1}{\gamma_{t}^{2}+\vect{p}^{2}}\left(\kappa_{1}g^{{}^{3}\!S_{1}-{}^{1}\!P_{1}}-2\kappa_{0}g^{{}^{3}\!S_{1}-{}^{3}\!P_{1}}\left(1+\frac{1}{\gamma_{t}}(\gamma_{t}-i|\vect{p}|)\right)\right)\frac{1}{2}(Z_{t}-1)\\\nonumber
&-L_{2}\sqrt{\frac{8\gamma_{t}}{M_{N}}}\frac{1}{\gamma_{t}+i\vect{p}}g^{{}^{3}\!S_{1}-{}^{3}\!P_{1}},
\end{align}
while in the resummed ERE
\begin{equation}
{U_{2}}_{\NLO}^{(R)}=-L_{2}\sqrt{\frac{8\gamma_{t}Z_{t}}{M_{N}}}\frac{1}{\gamma_{t}-\frac{1}{2}\rho_{t}(\vect{p}^{2}+\gamma_{t}^{2})+i\vect{p}}g^{{}^{3}\!S_{1}-{}^{3}\!P_{1}}.
\end{equation}
Finally the NLO $Z$-paramertrization  correction to $U_{3}$ is
\begin{align}
{U_{3}}^{(Z)}_{NLO} =& \sqrt{\frac{8\gamma_{t}}{M_{N}}}\frac{1}{\gamma_{t}^{2}+\vect{p}^{2}}\kappa_{0}g^{{}^{3}\!S_{1}-{}^{3}\!P_{1}}
\left(\frac{1}{2}+ \frac{1}{\gamma_{t}}(\gamma_{t}-i|\vect{p}|)\right)(Z_{t}-1) \\ \nonumber
&+L_{2}\sqrt{\frac{8\gamma_{t}}{M_{N}}}\frac{1}{\gamma_{t}+i\vect{p}}g^{{}^{3}\!S_{1}-{}^{3}\!P_{1}},
\end{align}
and in the resummed ERE
\begin{equation}
{U_{3}}^{(R)}_{NLO}=L_{2}\sqrt{\frac{8\gamma_{t}Z_{t}}{M_{N}}}\frac{1}{\gamma_{t}-\frac{1}{2}\rho_{t}(\vect{p}^{2}+\gamma_{t}^{2})+i\vect{p}}g^{{}^{3}\!S_{1}-{}^{3}\!P_{1}}.
\end{equation}
There are additional contributions from these diagrams proportional to $g^{{}^{3}\!S_{1}-{}^{1}\!P_{1}}$, which however do not contribute to the asymmetry to the order we are considering and which we therefore neglect.

\section{Parity-violating asymmetry and results}\label{sec:results}

The unpolarized $np$ capture  cross section up to NLO is given by
\begin{align}
\label{eq:cross}
&\sigma=\frac{2e^2}{|\vect{v}_{rel}|}\frac{1}{4\pi}\left[|Y_{\LO}|^{2}\frac{1}{M_{N}^{3}}(\vect{p}^{2}+\gamma_{t}^{2})^{3}+2\mathrm{Re}[Y_{\LO}^{*}Y_{\NLO}]\frac{1}{M_{N}^{3}}(\vect{p}^{2}+\gamma_{t}^{2})^{3}\right.\\\nonumber
&\hspace{2.5cm}\left.+|{E1_{v}}_{\LO}|^{2}\vect{p}^{2}\frac{1}{M_{N}}(\vect{p}^{2}+\gamma_{t}^{2})+2\mathrm{Re}[{E1_{v}}_{\LO}^{*}{E1_{v}}_{\NLO}]\vect{p}^{2}\frac{1}{M_{N}}(\vect{p}^{2}+\gamma_{t}^{2})\right].
\end{align}
This and the following expressions are valid for amplitudes in both the $Z$-parametrization and resummed ERE formalisms.  To obtain predictions for the cross section we fit the value of $L_{1}$ to the cold $np$ capture cross section of $\sigma^{expt}=334.2\pm.5$ mb at an incident neutron speed of $v=2200$ m/s, which yields a value of $L_{1}\simeq -6.90$~fm in the $Z$-parametrization, and $L_{1}\simeq -4.02$~fm in the resummed ERE.  The isoscalar M1 amplitude $X$ starts to contribute to the  unpolarized $np$ capture cross section only at $\mathrm{N}^{2}\mathrm{LO}$.  Using detailed balance, the total cross section for the time-reversed process $\gamma d\to np$ is related to that of $np$ capture by
\begin{equation}
\label{eq:crossreverse}
\sigma(\gamma d\to np)=\frac{2M_{N}(k_{0}-E_{d})}{3k_{0}^{2}}\sigma(np\to d\gamma),
\end{equation}
where $k_{0}\approx \frac{\vect{p}^2+\gamma_t^2}{M_N}$ is the photon energy and $E_{d}=2.224575$~MeV is the deuteron binding energy.  

Next we consider the asymmetry $A_{L}^{\gamma}$ which is the asymmetry in the break-up cross sections with positive and negative helicity photons $\sigma_{+}$ and $\sigma_{-}$ in $\vec{\gamma}d\to np$ and is defined as
\begin{equation}\label{eq:Pgamma}
A_{L}^{\gamma}=\frac{\sigma_{+}-\sigma_{-}}{\sigma_{+}+\sigma_{-}}.
\end{equation}
The asymmetry results from an interference between the PC amplitude $Y$ and the PV amplitude $V$, the PC $E1_{v}$ and PV $U_{1}$, $U_{2}$, and $U_{3}$ amplitudes, as well as the PC $X$ and PV $W$ amplitudes, and up to NLO is given by
\begin{align}
\label{eq:circularpolarization}
&A_{L}^{\gamma}=-\frac{2}{k_{0}}\frac{1}{|Y_{\LO}|^{2}+|{E1_{v}}_{\LO}|^{2}\frac{\vect{p}^{2}}{k_{0}^{2}}}\Bigg[\mathrm{Re}[Y_{\LO}^{*}V_{\LO}+Y_{\NLO}^{*}V_{\LO}+Y_{\LO}^{*}V_{\NLO}]\\\nonumber
&-\left(\mathrm{Re}[Y_{\LO}^{*}V_{\LO}]-\frac{1}{3}\vect{p}^{2}\mathrm{Re}[U_{LO}{E1_{v}}_{LO}^{*}]\right)\frac{2\mathrm{Re}[Y_{\LO}^{*}Y_{\NLO}]+2\mathrm{Re}[{E1_{v}}_{\LO}^{*}{E1_{v}}_{\NLO}]\frac{\vect{p}^{2}}{k_{0}^{2}}}{|Y_{\LO}|^{2}+|{E1_{v}}_{\LO}|^{2}\frac{\vect{p}^{2}}{k_{0}^{2}}}\\\nonumber
&+2\mathrm{Re}[X_{\NLO}^{*}W_{\LO}]-\frac{1}{3}\vect{p}^{2}\mathrm{Re}[U_{\LO}{E1_{v}}_{LO}^{*}]-\frac{1}{3}\vect{p}^{2}\mathrm{Re}[U_{\NLO}{E1_{v}}_{LO}^{*}]-\frac{1}{3}\vect{p}^{2}\mathrm{Re}[U_{\LO}{E1_{v}}_{\NLO}^{*}]\vast],
\end{align}
where $U_{\LO}={U_{1}}_{\LO}+{U_{2}}_{\LO}+3{U_{3}}_{\LO}$ and $U_{\NLO}={U_{1}}_{\NLO}+{U_{2}}_{\NLO}+3{U_{3}}_{\NLO}$.

We fit the parameter $L_{2}$ occurring in $X$, $U_{2}$, and $U_{3}$ to reproduce the deuteron magnetic dipole moment which up to NLO in \EFT in the $Z$-parametrization is given by \footnote{For analogous expressions using different conventions see Refs.~\cite{Kaplan:1998sz,Chen:1999vd}.}
\begin{equation}
\mu_{M}^{(Z)}=\Big(2Z_{t}\kappa_{0}+2L_{2}\gamma_{t}\Big),
\end{equation}
and in the resummed ERE by
\begin{equation}
\mu_{M}^{(R)}=\Big(2Z_{t}\kappa_{0}+2L_{2}\gamma_{t}Z_{t}\Big),
\end{equation}
where $\mu_{M}=.85741$ is the deuteron magnetic dipole moment.  From the fit we find $L_{2}\simeq~-1.36$~fm in the $Z$-parametrization, and $L_{2}\simeq~-0.805$~fm in the resummed ERE. 

To obtain estimates for the PV LECs we relate them to the DDH parameters using the results in Ref.~\cite{Vanasse:2011nd}.  As discussed, no definite determination of these DDH parameters currently exists. 
We therefore use three different sets of values, which are labeled DDH, DDH-adj, and Bowman in the following. DDH represents the DDH ``best values'' \cite{Desplanques:1979hn}. 
DDH-adj refers to a set in which two combinations of $\rho$ and $\omega$ couplings are fit to data on the $\vec{p}p$ longitudinal asymmetry, while the remaining couplings take the DDH ``best values'' \cite{Schiavilla:2004wn}. 
The set labeled Bowman is obtained by fitting the PV couplings to a variety of available data \cite{Bowman:int07}. With these three sets, we obtain the NLO results in Fig.~\ref{fig:Pgamma}.  
The thin solid line is the NLO $A_{L}^{\gamma}$ using DDH ``best values" in the $Z$-parametrization and the thicker solid line is the same result in the resummed ERE.  
The long-dashed thin line is the NLO $A_{L}^{\gamma}$ using the DDH-adj values in the $Z$-parametrization and the thicker long-dashed line is the result in the resummed ERE.  
Finally, the small-dashed thin line is the NLO $A_{L}^{\gamma}$ in the $Z$-parametrization using the Bowman values and the small-dashed thick line is the same result, but using the resummed ERE.  
The difference between the $Z$-parametrization and resummed ERE results is at most approximately 10\% at threshold and this is in line with the 10\% error that we expect in the $Z$-parametrization at NLO.  
We also note that between the DDH ``best values" and the Bowman values there is about a factor of two difference.  
We want to stress that because of the uncertainties in the DDH values the predictions in the plots are merely representations of possible values for $A_{L}^{\gamma}$. 
Using the full DDH allowed ranges we find a large variation in the value of $A_{L}^{\gamma}$ over several orders of magnitude and different signs. 
In addition, as pointed out in Ref.~\cite{Schindler:2013yua}, the relations between LECs and model parameters contain sizable uncertainties as they can strongly depend on several regularization scales.  
Therefore, even with the LECs estimated using the DDH ``best values,'' a comparison with previous model calculations is not very reliable. 
Keeping these caveats in mind, we note that in using the matched LECs the magnitude of our results  
tends to be larger than the values found in model calculations.
For example, the asymmetries using the AV18 potential and DDH ``best values'' in Refs.~\cite{Liu:2004zm,Schiavilla:2004wn} are about an order of magnitude smaller. 
The difference is smaller when the Bonn potential is used \cite{Schiavilla:2004wn}. 
We also note that our results are in better agreement with those of Ref.~\cite{Khriplovich:2000mb}. However, as explained above, due to the uncertainties in matching the DDH parameters to the LECs as well as the strong dependence on the choice of PC interactions in the model calculations, such differences are not surprising.
Using power counting arguments, we stress that for a given set of LECs, we estimate the theoretical error of our calculation to be of the order of 10\%.
\begin{figure}[hbt]
\includegraphics[width=100mm,angle=-90]{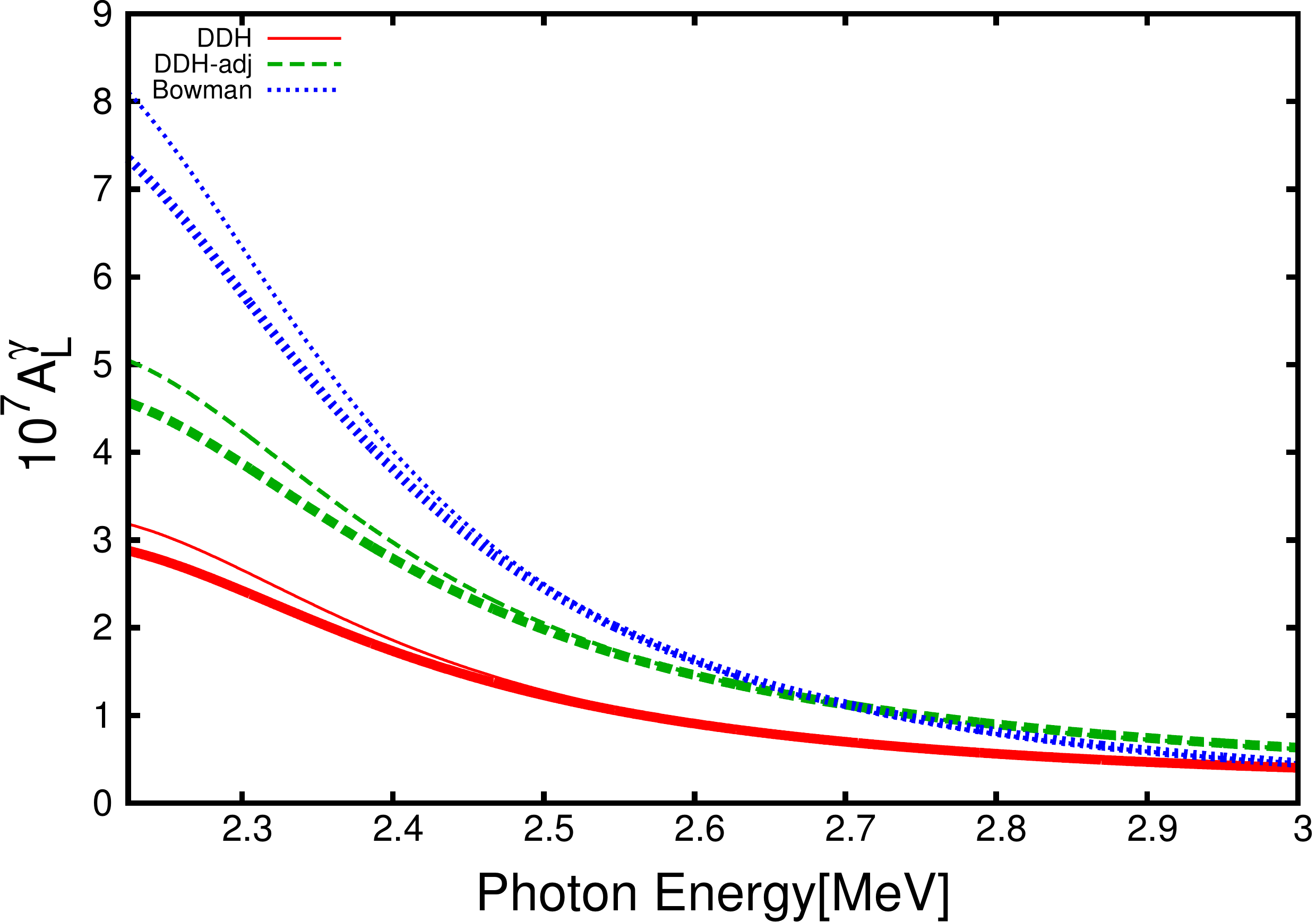}
\caption{\label{fig:Pgamma}(Color online) The PV asymmetry $A_{L}^{\gamma}$, as a function of the photon energy.  The lowest photon energy on the plot is the breakup threshold value corresponding to the deuteron binding energy.  The thin lines are $Z$-parametrization calculations and the thick lines resummed ERE calculations.}
\end{figure}

In addition to values based on the DDH estimates,  we also match the LECs to values of PV meson-nucleon couplings obtained from a nonlinear chiral Lagrangian in combination with a soliton model of the nucleon \cite{Kaiser:1989fd,Meissner:1998pu}. These values tend to be smaller than those from the DDH approach. The obtained asymmetry is also smaller than in the case of DDH-matched LECs as can be seen in Fig.~\ref{fig:PGammaSkyrme}. However, the values of $A_L^\gamma$ based on the soliton model are included in the variation of $A_L^\gamma$ when considering the complete DDH ranges.
\begin{figure}[hbt]
\includegraphics[width=100mm,angle=-90]{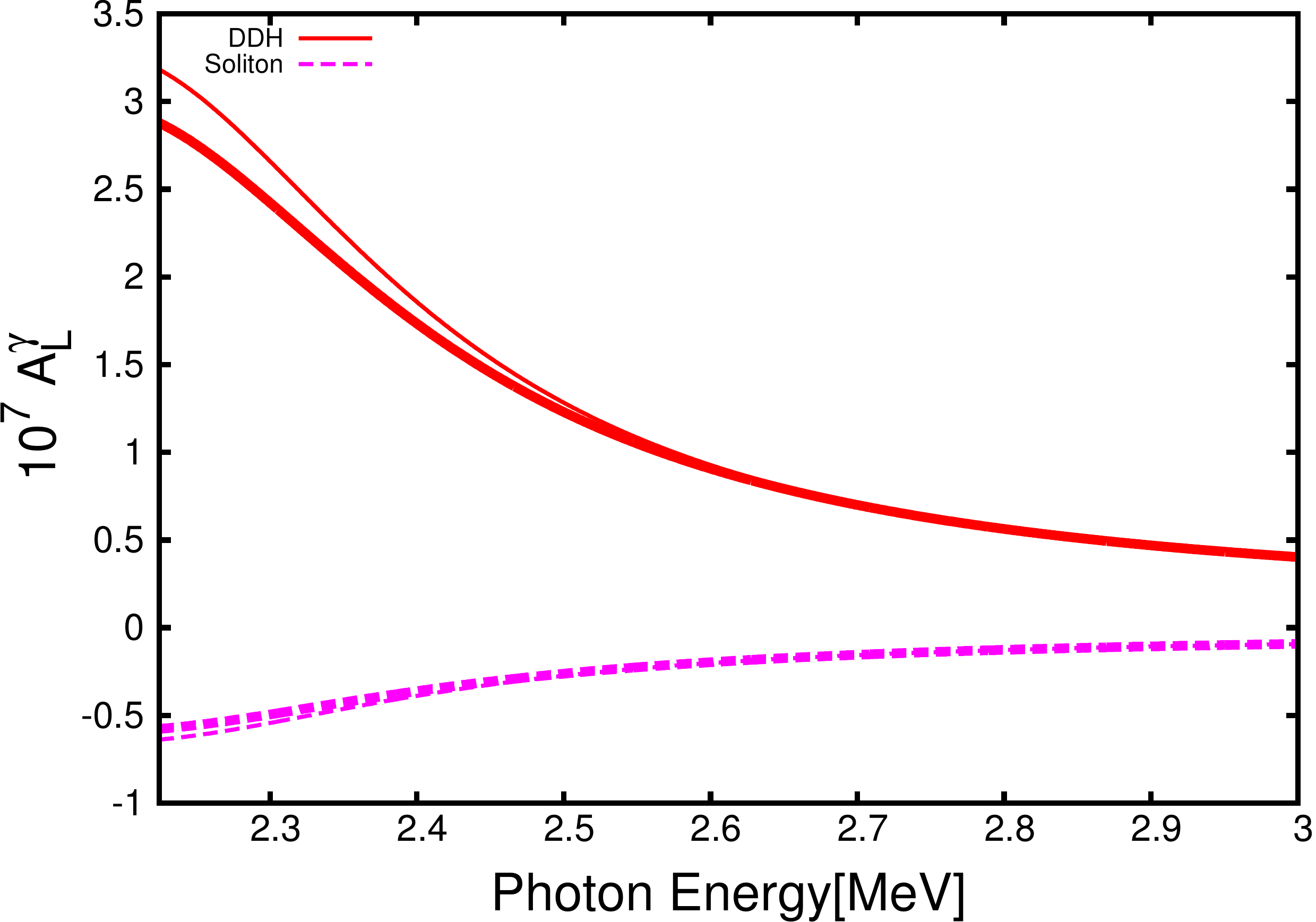}
\caption{\label{fig:PGammaSkyrme}(Color online) The PV asymmetry $A_{L}^{\gamma}$, as a function of the photon energy.  The lowest photon energy on the plot is the breakup threshold value corresponding to the deuteron binding energy.  The thin lines are $Z$-parametrization calculations and the thick lines resummed ERE calculations.}
\end{figure}

An alternative to determining the LECs that avoids model dependence is to estimate their size based on the assumption that they are ``natural,'' i.e., that the magnitude of the LECs is of order 1 in the correct units.
However, this does not determine the signs of the LECs nor the relative size of different LECs, and can therefore only be viewed as an order-of-magnitude estimate.
As discussed in Ref.~\cite{Griesshammer:2011md}, naturalness arguments lead to an estimate of 
\begin{equation}
\left\vert g^{(X-Y)} \right\vert \approx 10^{-10} \MeV^{-\frac{3}{2}}.
\end{equation}
Similarly, by comparison of the d\EFT calculation of $\vec{p}p$ scattering \cite{Schindler:2013yua} with the experimental result at 13.6 MeV \cite{Eversheim:1991tg}, the size of the LECs can be estimated to be\footnote{The asymmetry in $\vec{p}p$ scattering constrains one particular linear combination of LECs in d\EFT. Here, we consider this as an estimate of the size of individual LECs. However, it is possible that the individual LECs are larger and that a cancellation occurs for this particular linear combination.} 
\begin{equation}
\left\vert g^{(X-Y)} \right\vert \approx 10^{-11} \MeV^{-\frac{3}{2}}.
\end{equation}
The values obtained from matching the LECs to DDH model parameters agree with these estimates. 
In particular, considering the full ``reasonable ranges'' of the DDH couplings covers the range of LEC values one would obtain from naturalness arguments. 

Naively one would like to measure the asymmetry $A_{L}^{\gamma}$ at the photon momentum for which it is maximum.  
This value occurs at threshold for $\gamma d \to np$ where the cross section in the denominator of Eq.~\eqref{eq:Pgamma} is zero. However, this also means that the resulting count rate will be negligible.  
Therefore, it is important to find a photon energy for which the asymmetry is not too small, but the expected count rate is large enough to perform the experiment. 
A detailed determination of this energy would also include an analysis of systematic uncertainties and various issues of the actual construction of the experimental apparatus. 
Such an analysis is beyond the scope of this paper. Here, we merely try to find a suitable range of photon energies which balance the size of the asymmetry and the requirement for large enough count rates.  
To do so, we choose as a crude figure of merit $f=({A_{L}^{\gamma}})^{2}\times\sigma(\gamma d \to np)$.  
Using our results for $A_{L}^{\gamma}$ and again the DDH, DDH-adj, and Bowman values, as well as Eqs.~(\ref{eq:cross}) and (\ref{eq:crossreverse}) we find the figure of merit $f$ as is given in Fig.~\ref{fig:merit}.
\begin{figure}[h!bt]
\includegraphics[width=78mm,angle=-90]{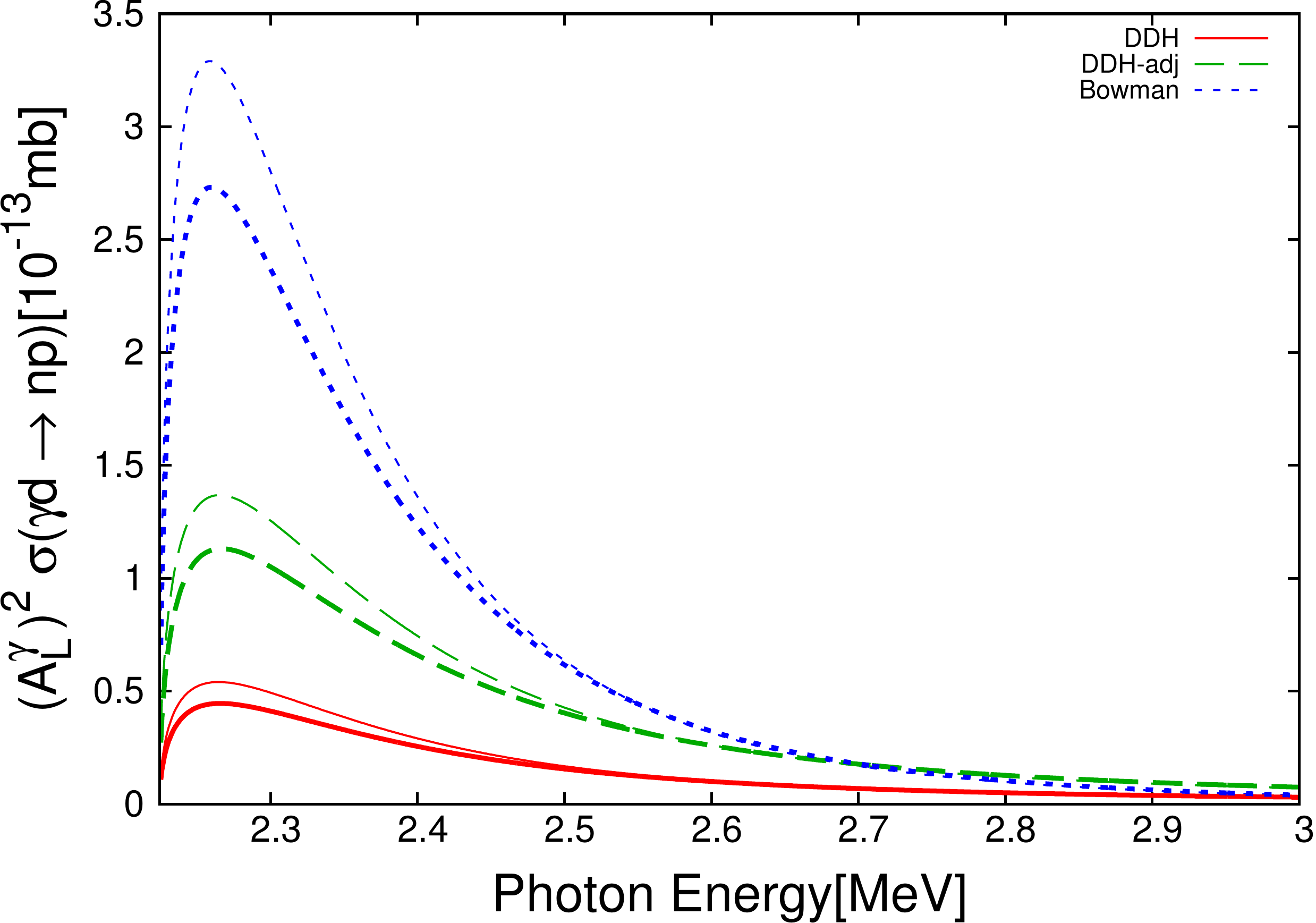}
\caption{\label{fig:merit}(Color online) The figure of merit $({A_{L}^{\gamma}})^{2}\times\sigma$ as a function of photon energy.  The lowest photon energy on the plot is the breakup threshold value corresponding to the deuteron binding energy.  The thin lines are $Z$-parametrization calculations and the thick lines resummed ERE calculations.}
\end{figure}
\begin{figure}[h!bt]
\includegraphics[width=78mm,angle=-90]{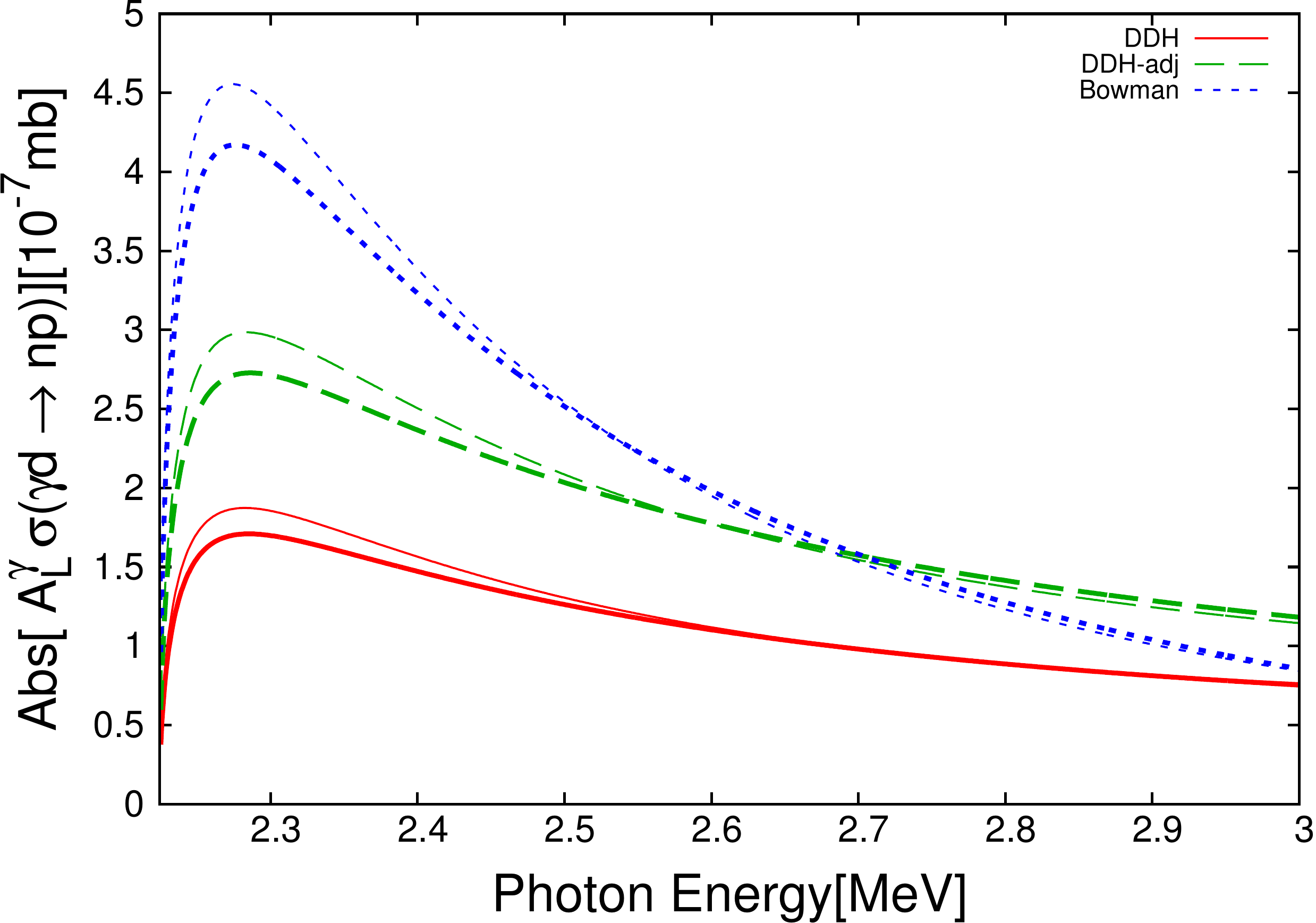}
\caption{\label{fig:merit2}(Color online) The figure of merit $A_{L}^{\gamma}\times\sigma$ as a function of photon energy.  The lowest photon energy on the plot is the breakup threshold value corresponding to the deuteron binding energy.  The thin lines are $Z$-parametrization calculations and the thick lines resummed ERE calculations.}
\end{figure}
The notation used for Fig.~\ref{fig:merit} is the same as in Fig.~\ref{fig:Pgamma}.  The figure of merit $f$ is maximized at a photon energy of $k=2.264$~MeV for both the $Z$ parametrization and resummed ERE using DDH and DDH-adj values.  
For the Bowman values, the figure of merit is maximized at a photon energy of $k=2.259$~MeV for both the $Z$-parametrization and the resummed ERE formalism.

As an additional estimate, we also consider the product $A_{L}^{\gamma}\times\sigma$ as an alternative figure of merit. As shown in Fig.~\ref{fig:merit2} and expected from the energy dependence of the asymmetry and the cross section, the maxima for this function correspond to slightly higher photon energies than those of Fig.~\ref{fig:merit}.  The figures of merit using the soliton LECs show maxima for similar photon energies.  While these are only very rough approximations to more reliable figures of merit, their combination suggests performing a measurement of the asymmetry around a photon energy of approximately 2.26 MeV to 2.3 MeV.

\section{Conclusion}\label{sec:concl}

We have calculated the energy dependence of the PV asymmetry $A^\gamma_L$ for deuteron breakup with polarized photons in \EFT to NLO, in which PC and PV interactions were treated within a consistent framework.
Based on power counting arguments, the theoretical errors in our results were estimated to be about 10\%. 
Using various estimates for the PV LECs, which have not been reliably determined at this stage, the asymmetry $A^\gamma_L$ is expected to be of the order of $10^{-7}$.
However, considering the large ``reasonable ranges'' for PV couplings that have been given in Ref.~\cite{Desplanques:1979hn}, variations of $A^\gamma_L$ of several orders of magnitude are possible.

With the continuing development of high-intensity photon sources and improvements in the required high control of systematics, a measurement of $A^\gamma_L$ might be possible in the future, and is currently being explored as a possibility at an upgraded HI$\gamma$S.
While a detailed analysis of possible sources of false asymmetries and of the experimental set-up is beyond the scope of this paper, we considered two simplified figures of merit to determine a range of photon energies that is suitable to balance the expected size of the asymmetry with the expected count rates. 
We found that measurements between 2.26 MeV and 2.3 MeV might be best suited for an experimental determination of $A^\gamma_L$.

Together with the existing measurement of the longitudinal asymmetry in $\vec{p}p$ scattering \cite{Eversheim:1991tg} and the expected results from the photon asymmetry in polarized neutron capture on protons, $\vec{n} p \to d \gamma$, \cite{Gericke:2011zz} such a measurement would provide important input into the determination of the PV coupling constants. 
The restriction to two-body systems should significantly reduce the theoretical uncertainties that are often present in the extraction of these couplings from systems involving larger numbers of nucleons.

\begin{acknowledgments}
We would like to thank Roger Carlini, Harald Grie{\ss}hammer, Mike Snow, Roxanne Springer, and other participants of the HIGS2 workshop for useful discussions.  We are also grateful to Roxanne Springer for comments on the manuscript.  This work is supported in part by the US Department of Energy under Grant no. DE-SC0010300 and Grant No. DE-FG02-05ER41368.
\end{acknowledgments}

\end{document}